\documentclass[a4paper]{article}

\usepackage{INTERSPEECH2019}
\usepackage{cite}
\usepackage{bm}
\usepackage{url}

\title{End-to-end Adaptation with Backpropagation through WFST\\
  for On-device Speech Recognition System}
\name{Emiru Tsunoo, Yosuke Kashiwagi, Satoshi Asakawa, Toshiyuki Kumakura}
\address{
  Sony Corporation, Japan}
\email{Emiru.Tsunoo@sony.com, Yosuke.Kashiwagi@sony.com, Satoshi.Asakawa@sony.com, Toshiyuki.Kumakura@sony.com}

\begin{document}

\maketitle
\begin{abstract}
An on-device DNN-HMM speech recognition system efficiently works with a limited vocabulary in the presence of a variety of predictable noise.
In such a case, vocabulary and environment adaptation is highly effective.
In this paper, we propose a novel method of end-to-end (E2E) adaptation, which adjusts not only an acoustic model (AM) but also a weighted finite-state transducer (WFST).
We convert a pretrained WFST to a trainable neural network and adapt the system to target environments/vocabulary by E2E joint training with an AM.
%This conversion is invertible; thus, the trained WFST parameters can be reconverted into an ordinary WFST and utilized in a conventional manner.
We replicate Viterbi decoding with forward--backward neural network computation, which is similar to recurrent neural networks (RNNs).
By pooling output score sequences, a vocabulary posterior for each utterance is obtained and used for discriminative loss computation.
Experiments using 2--10 hours of English/Japanese adaptation datasets indicate that the fine-tuning of only WFSTs and that of only AMs are both comparable to a state-of-the-art adaptation method, and E2E joint training of the two components achieves the best recognition performance.
We also adapt each language system to the other language using the adaptation data, and the results show that the proposed method also works well for language adaptations.
\end{abstract}

\section{Introduction}
Conventional automatic speech recognition (ASR) systems, typically referred to as DNN-HMM ASRs, commonly consist of DNN acoustic models (AMs), which estimate the posterior probabilities of hidden Markov model (HMM) states discriminatively, and language models (LMs), which are typically word $n$-grams \cite{bourlard94}.
Triphone HMMs and LMs are trained separately and incorporated together into a weighted finite-state transducer (WFST) for computation efficiency.
Further, in the case of on-device ASR systems, both AMs and LMs are compressed to reduce computational cost \cite{lei13,wang15}.
However, framewise cross-entropy training of AMs does not always minimize word error rates (WERs) because LMs model word sequences that have different granularities.
To fill the gap between the AMs and LMs, sequence discriminative training has been studied and has significantly reduced WERs \cite{povey05,kingsbury09,vesely13,su13,povey16}.

Recently, end-to-end (E2E) ASRs have attracted much attention as methods of directly integrating of AMs and LMs, including connectionist temporal classification (CTC) \cite{graves06, graves14, miao15}, attention-based encoder--decoder models \cite{chorowski15, chan16, lu16}, and hybrid models \cite{kim17, watanabe17}.
As the E2E ASRs directly map speech input frames into an output label sequence, the parameters are expected to be more optimized than those in conventional DNN-HMM ASRs in which modules are trained separately.
%However, most of E2E ASRs which beat conventional DNN-HMM ASR have a disadvantage that they require more than thousands of hours of speech-transcription parallel data \cite{prabhavalkar17,li17}.
In general, E2E ASRs, which are superior to conventional DNN-HMM ASRs, require thousands of hours of speech-transcription parallel data \cite{prabhavalkar17,li17}.

In the cases of ASR systems such as those in a robot, where most of the noise sounds are predictable, environment adaptation is highly effective.
Further, if the vocabulary is limited, better performance is expected using the adaptation data of such vocabulary.
%Instead of training E2E systems from scratch, we aim to exploit pre-trained models and use them as initial parameters for end-to-end re-training with small amount of data.
%Generally the small data is referred as adaptation data in a variety of studies which adapt models to speaker/environment/dictionary.
Among the adaptation methods, there are three major directions, input feature transformation \cite{neto95,yao12}, the use of auxiliary features \cite{saon13,delcroix16}, and the direct adaptation or transformation of DNN parameters~\cite{gemello06,yu13,swietojanski14,klejch18}, in which the last is related to this work the most.
However, all of them are concerned only with acoustic modeling.

In this paper, we propose a novel method of E2E adaptation.
We convert a pretrained WFST to a trainable neural network and adapt the system to the target environments/vocabulary by E2E joint retraining with the pretrained AM.
This conversion is invertible; thus, the trained WFST parameters can be reconverted into an ordinary WFST and utilized in a conventional manner.
We replicate Viterbi decoding with forward--backward neural network computation, which is similar to recurrent neural networks (RNNs).
For an application of on-device command recognition with a vocabulary size of hundreds of words, a vocabulary posterior for each utterance is obtained by simply pooling output score sequences, which is used for discriminative loss computation.
Experiments using 2--10 hours of English/Japanese adaptation data indicate that the fine-tuning of only the WFSTs and that of only the AMs are both comparable to a state-of-the-art adaptation method, and E2E joint training of the two components achieves the best recognition performance.
We also adapt each language system to the other language using the adaptation data, and the results show that the proposed method also works well for language adaptations.

The rest of the paper is organized as follows.
In Section \ref{sec:viterbinet} we describe our proposed WFST conversion algorithm to a trainable neural network.
In Section \ref{sec:impl} we briefly give some guidelines for an efficient implementation.
We describe our experiments in Section \ref{sec:experiment} and conclude the paper in Section \ref{sec:conclusion}.

\section{Backpropagation through WFST}
\label{sec:viterbinet}
\begin{figure}[t]
%\vspace{-3.5cm}
  %\centering
  \hspace{1.5cm}
  \includegraphics[width=0.8\columnwidth]{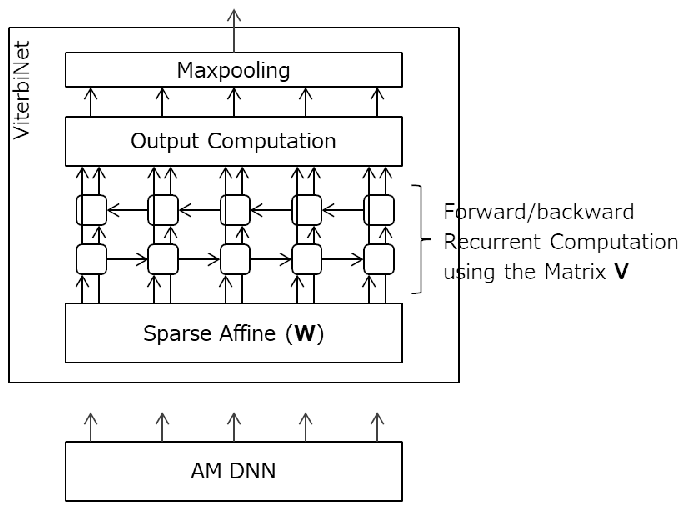}
  \caption{ViterbiNet: Neural network structure for Viterbi decoding.}
  \label{fig:viterbinet}
\end{figure}

To exploit pretrained AMs and LMs, and to improve the performance with a small amount of adaptation data, we convert a WFST to make it jointly trainable.
Although sequence discriminative training optimizes the AM training in a joint criterion with a given WFST \cite{povey05,kingsbury09,vesely13,su13,povey16}, to the best of our knowledge, no previous study has been carried out on updating a WFST during the training process, because it is complicated in general and the update might interfere with the LM.
Since Viterbi decoding is a simple linear procedure, it is differentiable.
Therefore, we alternate Viterbi decoding with a combination of neural network modules, and we see a DNN AM and the neural network decoder based on a WFST as a joint E2E system that is to be optimized.
The structure of the neural network is shown in Figure~\ref{fig:viterbinet} and we refer to this network as ViterbiNet in this paper.

\subsection{Forward--backward Computation}
First, each AM posterior is mapped to WFST states associated with the posterior by a sparse affine transformation.
This is equivalent to the transition model mapping followed by ilabel (input id of arc) selection for each state in Kaldi terminology.
Denoting $\mathbf{x}^{(t)}\in \mathbb{R}^P$ as the $P$-dimensional posterior of time frame $t$ (an output of the AM neural network), the transformation can be described as
\begin{equation}
  \mathbf{\tilde{x}}^{(t)} = \mathbf{W}\mathbf{x}^{(t)},
  \label{eq:mapping}
\end{equation}
where each element of $\mathbf{W}\in \mathbb{R}^{S\times P}$ is defined as 
\begin{equation}
  w_{i,j} = \begin{cases}
    1 & \text{(TransitionIdToPdf}(j) = \text{ilabel}(i)) \\
    0 & \text{(otherwise)}
  \end{cases}
  \label{eq:mapping_elements}
\end{equation}
and $S$ is the total state number.
The matrix $\mathbf{W}$ is constant and kept unchanged in this paper.

\begin{figure}[t]
%\vspace{-3.5cm}
  \centering
  %\hspace{-1.3cm}
  \includegraphics[width=1.0\columnwidth]{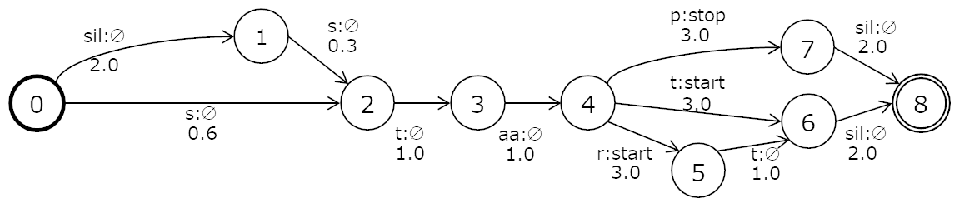} \\
  {\tiny
  \begin{equation*}
    \bm V=\left(
    \begin{array}{ccccccccc}
      0.0 \\
      2.0 & 0.0 \\
      0.6 & 0.3 & 0.0 \\
      & & 1.0 & 0.0 \\
      & & & 1.0 & 0.0 \\
      & & & & 3.0 & 0.0 \\
      & & & & 3.0 & 1.0 & 0.0 \\
      & & & & 3.0 & & & 0.0  \\
      & & & & & & 2.0 & 2.0 & 0.0 \\ 
    \end{array}
    \right)
  \end{equation*}
  }
  \caption{Example of sparse matrix representation of WFST.}
  \label{fig:wfst}
\end{figure}

The WFST transition can be represented as a sparse matrix $\mathbf{V} \in \mathbb{R}^{S \times S}$.
An example of a WFST is illustrated in Figure \ref{fig:wfst}, where the triphone HMM transitions and self-loops are omitted for simplicity.
In this case, the graph is represented as $\mathbf{V}$ in Figure \ref{fig:wfst}.
Let the forward score for each state at time frame $t$ be $\bm\alpha^{(t)} \in \mathbb{R}^{S}$.
This can be computed with matrix $\mathbf{V}$ as
\begin{align}
  \bm\alpha^{(t)} &= f(\mathbf{\tilde{x}}^{(t)}, \bm\alpha^{(t-1)}) \nonumber \\
  &= \mathbf{V} \otimes \left( \bm\alpha^{(t-1)} \circ \mathbf{\tilde{x}}^{(t)} \right),
  \label{eq:forward}
\end{align}
where $\circ$ is element-wise multiplication and $\otimes$ is the maximum value in each row after multiplying the elements of $\mathbf{V}$ by those of $\bm\alpha^{(t-1)} \circ \mathbf{\tilde{x}}^{(t)}$.
Thus, $\bm\alpha^{(t)}$ can also be written as 
\begin{equation}
  \alpha^{(t)}_{i} = \max_{j} v_{i,j} \alpha^{(t-1)}_{j} \tilde{x}^{(t)}_{j}.
  \label{eq:viterbi}
\end{equation}
Typically, the initial state of a WFST is set to $0$; therefore, the initial score $\bm\alpha^{(0)}$ is
\begin{equation}
  \alpha^{(0)}_{i} = \begin{cases}
    1 & (i=0) \\
    0 & \text{(otherwise)}.
  \end{cases}
  \label{eq:init}
\end{equation}
The above manipulation is mathematically equivalent to Viterbi decoding, and this can be carried out recursively, similarly to RNNs.

In general, to decode the output sequence, all olabels (output ids of arcs) of highest-scoring forward paths have to be memorized.
Instead, in this paper, we also compute the backward score, similarly to (\ref{eq:forward}), as
\begin{equation}
  \bm\beta^{(t)} = \left( \mathbf{V}^{\mathsf{T}} \otimes \bm\beta^{(t+1)} \right) \circ \mathbf{\tilde{x}}^{(t+1)}.
  \label{eq:backward}
\end{equation}
The final score of the WFST, $\bm\beta^{(T)}$, can be defined using the final score of each state.
By combining forward and backward computations, the scores of output symbols are calculated, which is described in the following section.

\subsection{Output Layer}
\label{sec:out}
Each arc between states has an olabel, and if the olabel is not $<$eps$>$, then an output symbol is emitted at each time frame.
Therefore, the output score $\mathbf{y}^{(t)}\in \mathbb{R}^{C}$ with vocabulary size $C$ is defined as
\begin{equation}
  y^{(t)}_{u} = \max_{v_{i,j}\in \mathbb{U}}\alpha^{(t-1)}_{i} v_{i,j} \tilde{x}_{i}^{(t)}\beta^{(t)}_{j},
  \label{eq:output}
\end{equation}
where $\mathbb{U}$ is a set of arcs whose olabels are $u$.
This resembles to Baum--Welch algorithm, except it takes maximum instead of summation for decoding.
Then, the score sequence $\mathbf{y}$ can be aligned with the reference word sequence using a similar method to the forward--backward algorithm in CTC training \cite{graves12}.  % by maximizing
%\begin{equation}
%  J = \sum_{r=1}^{R} \max_{t_{(r-1)} < t < t_{(r+1)}} y_{\text{ref}(r)},
%  \label{eq:object}
%\end{equation}
%where $t_{(r)}$ is the frame number aligned to the $r$-th reference word, and $\text{ref}(r)$ is the $r$th reference word id.
However, in this paper, only a small vocabulary size is used; therefore, it is feasible to concatenate all words of each command and represent them as single words (e.g., ``Take-a-picture'').
Taking this into account, the score of the output command $u$ can be simply written as
\begin{equation}
  \hat{y}_{u} = \max_{t}y_{u}^{(t)},
  \label{eq:command_output}
\end{equation}
which is realized with maxpooling manipulation.

\subsection{Loss Function}
The decoding is simplified to a command classification using the output score $\mathbf{\hat{y}}$.
The loss function is 
\begin{align}
  L &= \text{CrossEntropy}(\frac{\mathbf{\hat{y}}}{\lVert\mathbf{\hat{y}}\rVert_{1}},u) \nonumber \\
  &= \text{SoftMaxCrossEntropy}(\log\mathbf{\hat{y}},u),
  \label{eq:loss}
\end{align}
where $\lVert\cdot\rVert_{1}$ is the $l_{1}$-norm of the vector.
Since all aforementioned transformations including affine, forward--backward, and output computations are linear manipulations, the gradients for backpropagation are straightforwardly derived.

If only the AM is updated with the gradients, it acts similarly to sequence discriminative training~\cite{povey05,kingsbury09,vesely13,su13,povey16}.
The E2E training is carried out by updating both the AM and the matrix $\mathbf{V}$, which is the WFST itself.
In the case of a simplified single-word model, the loss function becomes exactly equivalent to the MMI criterion \cite{povey05}.

\section{Implementation}
\label{sec:impl}
We have theoretically discussed the proposed method.
To realize it in feasible manner, several guidelines are given.

\subsection{WFST Preprocess}
The WFST parameters are converted to a sparse matrix $\mathbf{V}$ and applied in every time step.
In general, however, the WFST contains a number of transitions that do not require any emissions, i.e., epsilon arcs.
As long as the WFST contains such nonemitting transitions, the proposed forward--backward procedure cannot be realized with the aforementioned recursive scheme.
Therefore, the WFST should first be processed with \texttt{fstrmepsilon} to remove epsilon arcs as much as possible.
In many cases, not all the epsilons are removed but the number becomes much smaller (e.g., 0.036\% of arcs for Switchboard trigram LM).
The surviving epsilon arcs are omitted from the matrix $\mathbf{V}$ for training and only utilized for evaluation in this paper.

\subsection{Sparse Computation}
The mapping transform matrix $\mathbf{W}$ in (\ref{eq:mapping}) and the transition matrix $\mathbf{V}$ in (\ref{eq:forward}) are sparse matrices.
The use of ordinary affine transform modules is impractical, especially for large WFST parameters.
Therefore, we encode the sparse values in the matrices and compute only these elements.
This is inefficient and memory-consuming when carried out using GPUs.
We leave this computation to the CPU.

\subsection{Log-scale Computation}
The proposed forward--backward computation also requires a series of multiplications over the input utterance.
This leads to rounding errors in computation.
Although normalization of the values in each step can improve the precision, we adopt log-scale computation for the recursive multiplication similarly to Viterbi decoding.
In this case, the multiplications in (\ref{eq:forward}) and (\ref{eq:backward}) turn into summations of logarithms, and the gradients of the forward--backward layer become constant values delivered from the output layer and spread along the possible transitions.
Therefore, the log-scale computation increases the speed of inference as well as backpropagation.

\subsection{Regularization of AM}
Sequence discriminative training tends to overfit \cite{su13} and the proposed E2E training is not an exception.
Since we exploit the pretrained AM, we adopt Kullback--Leibler (KL) divergence regularization, as in \cite{yu13}.
By adding this divergence as a regularization term to (\ref{eq:loss}), the optimization criterion becomes
\begin{equation}
  L' = \text{SoftMaxCrossEntropy}(\log\mathbf{\hat{y}},\hat{r}) + \lambda \sum_{t=1}^{T}\text{KL}(\mathbf{x}^{(t)}_{\text{org}} || \mathbf{x}^{(t)}),
  \label{eq:klreg}
\end{equation}
where $\mathbf{x}^{(t)}_{\text{org}}$ is the AM posterior of time frame $t$ computed with the original pretrained parameters, and $\lambda$ is the controllable regularization weight.
The regularization allows not only the AM to remain close to the original AM, but also the fast convergence because it spreads peaky gradients of the ViterbiNet over the dimensions of the posterior that represent similar HMM states.
%When we conducted a preliminary experiment on updating only AM parameters and changing the vocabulary of LM when decoding, we confirmed the robustness against overfitting to the LM and obtained some improvement.
The regularization of WFST parameters is left as future work.

\section{Experiments}
\label{sec:experiment}
We evaluated command recognition tasks with small-footprint setups for on-device systems such as those in robots.
US English and Japanese systems were built for the evaluation.
Three tasks were evaluated; English Speech Commands (EN-SC, 20 commands) \cite{speechcommands}, English robot commands (EN-ROBOT, 157 commands), and Japanese robot commands (JP-ROBOT, 229 commands).
The proposed method was compared with state-of-the-art sequence discriminative training and adaptation methods.
We also conducted language adaptations from Japanese to English and from English to Japanese using the same task setups.

\begin{table}[t]
  \caption{Adaptation and evaluation datasets for command recognition tasks}
  \label{tab:data}
  \centering
  {\footnotesize
  \begin{tabular}{l|ccc}
    \hline
    Task & EN-SC & EN-ROBOT & JP-ROBOT \\
    \hline
    \hline
    examples & ``cat'' & ``come-here'' & ``oide'' \\
    & ``three'' & ``move-forward'' & ``zenshin'' \\
    \hline
    Adaptation data \\
    vocabulary & 20 & 157 & 229 \\
    \# of utterances & 34,760 (9.6 h) & 3,297 (1.8 h) & 4,580 (2.1 h) \\
    \# of speakers & 1,034 & 7 & 20 \\
    SNR & 0 -- 20 dB & -5 -- 15 dB & -5 -- 15 dB \\
    \hline
    Evaluation data \\
    vocabulary & 20 & 157 & 229 \\    
    \# of utterances & 29,961 (8.3 h) & 1,413 (0.7 h) & 2,290 (1.0 h)\\
    \# of speakers & 847 & 3 & 10 \\
    SNR & 0 -- 20 dB & -5 -- 15 dB & -5 -- 15 dB \\
    \hline
  \end{tabular}
  }
\end{table}

\subsection{Experimental Setup}
The input speech signals were 16 kHz, 16-bit monaural speech data processed by multi-scale multi-band DenseNet-based noise reduction (NR) described in \cite{takahashi17}, which was trained using WSJ, TIMIT, and ATR-503 Japanese data.
The feature vector for every 10 ms was a 24-dimensional log filterbank and energy with $\Delta$ and $\Delta \Delta$ (75 dimensions in total).
The features were spliced into 11 frames (5 on each side of the current frame) followed by a 5-layer full connection DNN with 640 hidden units for each layer.
We built two AM DNNs, for English and Japanese, where the outputs of the AM DNNs were 1,256 units for the English model, and 1,012 units for the Japanese model.
The AMs were first trained solely with 28,000 hours of English data and 12,300 hours of Japanese data, respectively, which were mixed with mechanical and living noises, followed by fine-tuning with a 280 hour English subset and a 165 hour Japanese subset jointly with the front-end DenseNet NR.
We used these AMs for the baseline systems.

LMs were built for each task.
For the EN-SC task, a grammar-based WFST with a vocabulary of 20 words was built, which had 493 states and 1,452 arcs.
For the EN-ROBOT task, we constructed a grammar-based WFST of 225 commands, including all commands in the adaptation and evaluation sets (8,351 states, 21,879 arcs), and a grammar WFST of 305 commands (3,797 states, 8,014 arcs), including all adaptation/evaluation commands for the JP-ROBOT task.

For adaptation, we converted the WFSTs to ViterbiNets and jointly retrained them with the AMs using adaptation datasets (ViterbiNet E2E), specifically, by updating all parameters in the 5-layer AM DNN and the matrix $\mathbf{V}$.
The proposed network was implemented with the open source {\it Sony Neural Network Libraries}\footnote{Neural Network Libraries by Sony: \url{https://nnabla.org/}}.
Training was carried out using the Adam optimizer (learning rate $=0.0001$, $\beta_{1}=0.9$, $\beta_{2}=0.999$) with a 16-utterance minibatch, and 20 epochs were sufficient for convergence in every task.
The regularization parameter $\lambda$ in (\ref{eq:klreg}) was set to $0.01$.
We also evaluated the cases where only the AMs were updated (ViterbiNet AM) in the proposed scheme, which is theoretically equivalent to sequence discriminative training, and only the ViterbiNets, i.e., the matrix $\mathbf{V}$, were updated (ViterbiNet WFST).
We also compared the above retraining with the simple retraining of only AMs using aligned labels (Cross Entropy), sMBR discriminative training \cite{gibson06}, and state-of-the-art KL-regularization adaptation \cite{yu13}, using the same adaptation dataset with 40 iterations for sMBR and 20 epochs for the other methods.
The adaptation datasets contained all commands in the evaluation sets, but all speakers were different; thus only the environment and vocabulary matched.

\begin{table}[t]
  \caption{Sentence error rates (SERs) of adaptation methods in command recognition tasks}
  \label{tab:adaptation}
  \centering
  \begin{tabular}{l|ccc}
    \hline
    Adaptation methods  & EN-SC & EN-ROBOT  & JP-ROBOT \\
    \hline\hline
    (No adaptation)  & 34.70 & 8.99 & 6.24 \\ %& 8.70\\
    Cross Entropy & 10.78 & 6.44 & 4.06 \\%& 6.58 \\
    sMBR \cite{gibson06}  & 10.83 & 6.65 & 3.14 \\ % & 6.37\\
    KL-regularization \cite{yu13} & 10.77 & 6.23 & 3.28 \\ % & 5.73\\
    ViterbiNet AM & 9.65 & 6.09 & 2.93 \\ % & 5.31\\
    ViterbiNet WFST & 13.08 & 3.89 & 3.06 \\ % & 3.61\\
    ViterbiNet E2E & {\bf 9.26} & {\bf 3.54}& {\bf 2.66} \\ %& {\bf 3.18}\\
    \hline
  \end{tabular}
\end{table}

In the evaluation step, trained ViterbiNet parameters were converted back to WFSTs, and Viterbi decoding was applied as in conventional ASR.
The beam size was set to $7$ and the acoustic scale was $0.07$.
To all the EN-SC adaptation data and evaluation data, we added randomly chosen noise in the dataset (e.g., running tap) with a uniform distribution between an SNR of 0 dB and 20 dB before processing with the fine-tuned DenseNet NR.
To all the adaptation and evaluation datasets for EN-ROBOT and JP-ROBOT, we randomly added mechanical and living noise (e.g., servomotor sounds, music) an SNR from -5 dB to 15 dB before the NR.
The details of the adaptation and evaluation datasets are given in Table \ref{tab:data}.

\subsection{Environment/Vocabulary Adaptation Result}
The baselines and all the obtained sentence error rates (SERs) of adaptation methods are shown in Table \ref{tab:adaptation}.
For sMBR, the lowest SERs among the iterations are shown.
Compared with the baselines, all adaptations performed better since the conditions of vocabulary and environments matched.
The EN-SC task performed badly because the AM was trained with completely different noise mixtures, and all the commands in the task were simple and short; therefore, it was difficult to distinguish them.
Moreover, the environments varied since the data was collected from a number of volunteers in various rooms with different microphones.
The results of the ViterbiNet AM were similar to those of sMBR or even better since they were equivalent to the conventional MMI training, not in the state level but in the word level.
%Surprisingly, the results of ViterbiNet WFST were as good as those of ViterbiNet AM.
Surprisingly, updating only ViterbiNet WFST also significantly reduced SERs.
We suppose that the trained WFST can model sequential information rather than static information, which worked well, especially for the vocabulary adaptations.
The joint E2E training of the AM and WFST using the proposed ViterbiNet consistently achieved the best performances.

\subsection{Language Adaptation}
Next, we evaluated language adaptation from Japanese to English using the EN-SC and EN-ROBOT tasks and from Japanese to English using the JP-ROBOT task.
For the English tasks we used the Japanese AM as a seed, and new LMs were compiled to match the AM using Japanese phones and HMM.
The lexicons were designed manually, such as {\it s e b N} for ``seven'' and {\it f r i:} for ``three,'' since Japanese does not have {\it v} and {\it th} pronunciations.
Similarly, we used the English AM as a seed for the Japanese task and compiled the LM using English phones and HMM.
Then we applied each method to adapt to the language of the test tasks using the aforementioned adaptation datasets.
The results are shown in Table~\ref{tab:lang-adapt}.
The original results in the first row were poor because of the language mismatches of AMs and the ill-designed LMs.
However, with adaptations, we obtained similar tendencies to the results in Table~\ref{tab:adaptation} and achieved the best performance using the proposed E2E adaptation.
The E2E adaptation was so strong that it approached the performances of language-matched training, and the results imply that a few hours of adaptation data are sufficient to build a system for a task as simple as command recognition tasks, even from a model of a different language.
We also tested the adaptations from random initial values for the AMs, but the training did not progress at all.

\begin{table}[t]
  \caption{SERs of adaptation methods in command recognition tasks using AMs of different languages as initial seeds}
  \label{tab:lang-adapt}
  \centering
  \begin{tabular}{l|ccc}
    \hline
    Adaptation methods  & JP AM$\to$ & JP AM$\to$ & EN AM$\to$ \\
      & EN-SC & EN-ROBOT & JP-ROBOT \\    
    \hline\hline
    (No adaptation) & 84.12 & 94.13 & 46.03 \\ 
    KL-regularization \cite{yu13} & 67.35 & 54.64 & 16.94 \\ % klreg 0.001
    ViterbiNet AM & 54.40 & 31.21 & 10.39 \\ % klreg 0.001
    ViterbiNet WFST & 47.24 & 49.54 & 25.68 \\
    ViterbiNet E2E & {\bf 27.64} & {\bf 13.09} & {\bf 7.64} \\ % klreg 0.0
    \hline
  \end{tabular}
\end{table}

\section{Conclusions and Future Work}
\label{sec:conclusion}
We have proposed a new end-to-end adaptation method by fine-tuning an AM as well as a WFST.
We converted the pretrained WFST to a trainable neural network and adapted the system to target environments/vocabulary.
We replicated Viterbi decoding with forward--backward neural network computation, which is similar to RNNs.
By pooling the output score sequences, a vocabulary posterior for each utterance was obtained and used for a discriminative loss computation.
Experiments using 2--10 hours of English/Japanese adaptation data indicated that the fine-tuning of only WFSTs and that of only AMs were both comparable to a state-of-the-art adaptation method, and the E2E joint training of the two components achieved the best recognition performances.
We also adapted one language system to the other language using a small amount of adaptation data, and the results showed that the proposed method also worked well for language adaptations.

Our ongoing work includes training larger WFSTs, such as Switchboard trigram LM, for LVCSR.
Also, the proposed training method can be applied to external LMs for E2E ASR systems using CTC/Attention, where the complexity should be lower and the training becomes more feasible, especially with character/subword outputs.
%\section{Acknowledgements}

\newpage

\bibliographystyle{IEEEtran}

\bibliography{mybib}

\end{document}